\documentclass[a4paper,11pt]{article}
\pdfoutput=1 

\usepackage{jheppub} 

\usepackage[T1]{fontenc} 

\title{\boldmath Loop Corrections in Double Field Theory:  Non-trivial Dilaton Potentials}

\author[a]{Songlin Lv,}
\author[a]{Houwen Wu,}
\author[a,b]{Haitang Yang}

\affiliation[a]{Center for theoretical physics,\\College of Physical Science and Technology,\\Sichuan University, Chengdu, 610064, China}
\affiliation[b]{Kavli Institute for Theoretical Physics China (KITPC), \\Chinese Academy of Sciences, \\Beijing 100080, P.R. China}

\emailAdd{2013222020005@stu.scu.edu.cn}
\emailAdd{2013222020003@stu.scu.edu.cn}
\emailAdd{hyanga@scu.edu.cn}

\abstract{It is believed that the invariance of the generalised diffeomorphisms prevents any non-trivial dilaton potential from double field theory. It is therefore difficult to include loop corrections in the formalism. We show that by redefining a non-local dilaton field, under strong constraint which is necessary to preserve the gauge invariance of double field theory, the theory does permit non-constant dilaton potentials and loop corrections. If the fields have dependence on only one single coordinate, the non-local dilaton is identical to the ordinary one with an additive constant.}

\begin{document}
\maketitle
\flushbottom

\section{Introduction}

Born from closed string field theory, double field theory (DFT) \cite{Siegel:1993xq,Hull:2009mi,Tseytlin:1990nb, Duff:1989tf}
is constructed by formally doubling all spacetime coordinates of the massless
sector of closed string spectrum. DFT  manifests an
$O\left(D,D\right)$ symmetry explicitly and the dual coordinates represent the conjugation of winding numbers. Formally, all components
of closed string fields are formulated by double coordinates $\phi_{I}\left(X^{M}\right)$,
where $X^{M}=\left(\tilde{x}_{i},x^{i}\right)$, $M=1,2,\ldots,2D$
and $i=1,2,\ldots,D$. In such a formalism, $x^{i}$ is the usual
coordinates and $\tilde{x}_{i}$ denotes the dual coordinates
of winding momentum. Good reviews of DFT are given by
\cite{Zwiebach:2011rg,Aldazabal:2013sca,Berman:2013eva,Hohm:2013bwa}.

To build an $O\left(D,D\right)$ invariant spacetime action, a generalized $O\left(D,D\right)$ metric is introduced

\begin{equation}
\mathcal{H}_{MN}=\left(\begin{array}{cc}
g^{ij} & -g^{ik}b_{kj}\\
b_{ik}g^{kj} & g_{ij}-b_{ik}g^{kl}b_{lj}
\end{array}\right),
\end{equation}

\noindent unifying the spacetime metric $g_{ij}$ and anti-symmetric
Kalb-Ramond field $b_{ij}$ altogether. In this metric, $M$ is an $O\left(D,D\right)$
index,  running from $1$ to $2D$. The $O\left(D,D\right)$ invariant
spacetime action is built by contraction of $O\left(D,D\right)$
indices,

\begin{eqnarray}
S & = & \int d^{D}xd^{D}\tilde{x}e^{-2d}\left(\frac{1}{8}\mathcal{H}^{MN}\partial_{M}\mathcal{H}^{KL}\partial_{N}\mathcal{H}_{KL}-\frac{1}{2}\mathcal{H}^{MN}\partial_{N}\mathcal{H}^{KL}\partial_{L}\mathcal{H}_{MK}\right.\nonumber \\
 &  & \left.-2\partial_{M}d\partial_{N}\mathcal{H}^{MN}+4\mathcal{H}^{MN}\partial_{M}d\partial_{N}d\right),
\end{eqnarray}

\noindent where $d$ is an $O\left(D,D\right)$ scalar dilaton, defined by the usual dilaton $\phi$ through $e^{-2d}=\sqrt{g}e^{-2\phi}$.
Therefore, DFT is an effective theory for three massless fields:
$D$ dimensional spacetime metric $g_{ij}$, the anti-symmetric Kalb-Ramond
field $b_{ij}$ and the scalar dilaton $\phi$.
If we compactify $d$ dimensions of $D=n+d$,
the continuous $O\left(D,D\right)$ group breaks to $O\left(n,n\right)\times O\left(d,d;Z\right)$,
where $O\left(n,n\right)$ is still a continuous group and $O\left(d,d;Z\right)$
is  T-duality in the compactified background. To be a consistent theory,
DFT is required to be invariant under the gauge transformations

\begin{eqnarray}
\delta_{\xi}\mathcal{H}^{MN} & = & \hat{\mathcal{L}}_{\xi}\mathcal{H}^{MN}\equiv\xi^{P}\partial_{P}\mathcal{H}^{MN}+\left(\partial^{M}\xi_{P}-\partial_{P}\xi^{M}\right)\mathcal{H}^{PN}+\left(\partial^{N}\xi_{P}-\partial_{P}\xi^{N}\right)\mathcal{H}^{MP},\nonumber \\
\delta d & = & \xi^{M}\partial_{M}d-\frac{1}{2}\partial_{M}\xi^{M},
\end{eqnarray}

\noindent where $\xi^{M}=\left(\tilde{\xi}_{i},\xi^{i}\right)$ and
$\hat{\mathcal{L}}_{\xi}$ is the ``generalized Lie derivatives''.
Since DFT is based on closed string theory, it must satisfy the level
matching condition: $L_{0}-\bar{L}_{0}=-p_{i}w^{i}=0$ for massless
states. In the language of DFT, the level matching condition is transformed
to the weak constraint: $\partial^{M}\partial_{M}A=0$ where $\partial_{M}=\left(\tilde{\partial}^{i},\partial_{i}\right)$
and $A$ stands for an arbitrary field. However, the weak constraint is
insufficient to guarantee the gauge invariance. In order to make  $\tilde{\partial}^{i}\partial_{i}\left(\delta_{\xi}\Phi\right)=
\tilde{\partial}^{i}\partial_{i}\left(\xi\cdot\Phi\right)=0$,
a  much stronger constraint is imposed:
$\partial^{M}\partial_{M}\left(\cdot\right)$
where $\cdot$ denotes any product of fields or gauge parameters.
Under this strong constraint, only half of the coordinates survive for all fields and gauge parameters and DFT reduces to the $D$ dimensional traditional low energy effective theory. 
In addition, there are many
works to imply that the strong constraint can be relaxed on a torus
background, massive type IIA and gauged supergravity \cite{Hohm:2011cp,Aldazabal:2011nj,Geissbuhler:2011mx,Geissbuhler:2013uka,Berman:2013uda}.
The detailed discussions on constraint relaxation are
summarized in \cite{Hohm:2013bwa,Grana:2012rr}.

It is widely believed that a non-trivial dilaton
potential is forbidden by the generalised diffeomorphism in DFT \cite{Jeon:2011cn}. Considering the DFT action $S=\int dxd\tilde{x}e^{-2d}\mathcal{R}$,
it proves that $\mathcal{R}$ is an $O\left(D,D\right)$ scalar
and also a gauge scalar.  Moreover, since the weight of the $O(D,D)$ scalar $e^{-2d}$
equals the unity, it is a density which is invariant
under the generalised diffeomorphisms when combined with the proper volume $\int dx d\tilde x$.  The point is that  $\int dx d\tilde x e^{-2d}$ is the unique multiplying factor of the dilaton that respects the generalised diffeomorphisms and $O(D,D)$ symmetry. This is such a strong constraint that  higher loop corrections are completely excluded! It is worth noting that the dilaton will always increase
as time goes by and it also marks the growth of the curvature. However, the growth of the string coupling $g_{s}=\exp\left(2\phi\right)$
and the growth of the Hubble parameter $H$ lead the universe approaching
two limits \cite{Gasperini:2007zz}, or two corrections to the low energy effective action:
(1) the string curvature scale, which requires the $\alpha^{\prime}$
corrections to the low energy effective action when $\sqrt{2\pi\alpha^{\prime}} H$
reaches $1$, and (2) the strong coupling regime, which requires the
quantum loop corrections of the form $e^{2n\, \phi}\,(\cdots)$ for non-negative integer $n$, when $g_{s}\sim 1$.
The first expansion has been discussed in ref. \cite{Hohm:2013jaa}. It would be unnatural that loop corrections  totally disappear in the formalism,  since beyond both limits,
the universe enters the string non-perturbative regime, and the action
with these two corrections will give us some non-perturbative signatures
which should be described by the yet-to-know M-theory.

The aim of this paper is to address the higher loop quantum corrections in DFT. To achieve this purpose, one does not really need to consider the complete loop expansion but justification of a pure dilaton potential is sufficient. However, a simple dilaton  potential of the form $V\left(e^{d}\right)$ does not work, since the product $e^{-2d}V\left(e^{d}\right)$ is no longer a density  and breaks the gauge invariance of the action,  though it is an $O\left(D,D\right)$ scalar. It turns out that in order to preserve the symmetries, a non-local dilaton $d_{*}$ has to be defined to replace the ordinary local dilaton by the similar method used in the traditional string cosmology \cite{Gasperini:2003pb}. Since this non-local dilaton $e^{-2d_{*}}$ includes a proper volume, it does not break the gauge invariance under the strong constraint.  Then
additional potential terms  of DFT action could be any regular function of this non-local dilaton.  Moreover, if we choose the cosmological background and
the cosmic-time gauge, the non-local dilaton $e^{-2d_{*}}$ reduces
to the usual $O\left(D,D\right)$ scalar dilaton $e^{-2d}$ multiplied by a proper volume.
The cosmological implications with this non-local dilaton have been
discussed in our previous works \cite{Wu:2013sha,Wu:2013ixa}.
Ref. \cite{Ma:2014ala} discussed cosmological solutions
with a constant dilaton potential.

The reminder of this paper is outlined as follows. In section 2, we
discuss the gauge transformations of DFT. We define the
non-local dilaton in  Section 4 . Section 5 is our conclusion and discussions.

\section{Generalized Lie derivatives and gauge scalar}

To begin with, we give a brief review of the gauge transformations based on
refs. \cite{Hull:2009mi}.
The DFT action,  expanded  in terms of $g_{ij}$,
$b_{ij}$ and $d$, can be recasted as

\begin{equation}
S=S^{\left(0\right)}+S^{\left(1\right)}+S^{\left(2\right)},
\end{equation}

\noindent with

\noindent
\begin{equation}
S^{\left(k\right)}=\int dxd\tilde{x}\mathcal{L}^{\left(k\right)},\qquad k=0,1,2,
\end{equation}

\noindent where the superscript denotes the number of $\tilde{\partial}$
in the DFT action. The full gauge transformations can
be written as

\begin{eqnarray}
\delta_{\xi}g_{ij} & = & \mathcal{L}_{\xi}g_{ij}+\mathcal{L}_{\tilde{\xi}}g_{ij}+2\left(\tilde{\partial}^{k}\xi^{l}-\tilde{\partial}^{l}\xi^{k}\right)\left(g_{ki}b_{jl}+g_{kj}b_{il}\right),\nonumber \\
\delta_{\xi}g^{ij} & = & \mathcal{L}_{\xi}g^{ij}+\mathcal{L}_{\tilde{\xi}}g^{ij}-\left[\left(\tilde{\partial}^{i}\xi^{k}-\tilde{\partial}^{k}\xi^{i}\right)g^{jl}b_{lk}+\left(i\leftrightarrow j\right)\right],\nonumber \\
\delta_{\xi}b_{ij} & = & \mathcal{L}_{\xi}b_{ij}+\mathcal{L}_{\tilde{\xi}}b_{ij}+\partial_{i}\tilde{\xi}_{j}-\partial_{j}\tilde{\xi}_{i}+g_{ik}\left(\tilde{\partial}^{l}\xi^{k}-\tilde{\partial}^{k}\xi^{l}\right)g_{lj}+b_{ik}\left(\tilde{\partial}^{l}\xi^{k}-\tilde{\partial}^{k}\xi^{l}\right)b_{lj},\nonumber \\
\delta_{\xi}d & = & \left(\xi^{i}\partial_{i}+\tilde{\xi}_{i}\tilde{\partial}^{i}\right)d-\frac{1}{2}\left(\partial_{i}\xi^{i}+\tilde{\partial}^{i}\tilde{\xi}_{i}\right),
\end{eqnarray}

\noindent where $\mathcal{L}_{\xi}$ is the Lie derivatives with respect
to $\xi$, $\mathcal{L}_{\tilde{\xi}}$ is the dual Lie derivatives
with respect to $\tilde{\xi}$, $\xi$ and $\tilde{\xi}$ are gauge
parameters. The Lie derivatives and its dual for arbitrary tensors
$u_{i}^{\; j}$ can be defined as follows

\begin{eqnarray}
\mathcal{L}_{\xi}u_{i}^{\; j} & = & \xi^{p}\partial_{p}u_{i}^{\; j}+\partial_{i}\xi^{p}u_{p}^{\; j}+\partial_{p}\xi^{j}u_{i}^{\; p},\nonumber \\
\mathcal{L}_{\tilde{\xi}}u_{i}^{\; j} & = & \tilde{\xi}_{p}\tilde{\partial}^{p}u_{i}^{\; j}-\tilde{\partial}^{j}\tilde{\xi}_{p}u_{i}^{\; p}-\tilde{\partial}^{p}\tilde{\xi}_{i}u_{p}^{\; j}.
\end{eqnarray}

\noindent We can split it into two parts according to the number of $\tilde\partial$ acted

\noindent
\begin{equation}
\delta_{\xi}=\delta_{\xi}^{\left(0\right)}+\delta_{\xi}^{\left(1\right)}.
\end{equation}

\noindent These gauge transformations $\delta_{\xi}^{\left(0\right)}$
and $\delta_{\xi}^{\left(1\right)}$ are T-dual with each other. For example,
to consider the gauge transformations for dilaton $\delta_{\xi}d$, we have

\begin{eqnarray}
\delta_{\xi}^{\left(0\right)}d & = & \xi^{i}\partial_{i}d-\frac{1}{2}\partial_{i}\xi^{i},\nonumber \\
\delta_{\xi}^{\left(1\right)}d & = & \tilde{\xi}_{i}\tilde{\partial}^{i}d-\frac{1}{2}\tilde{\partial}^{i}\tilde{\xi}_{i}.
\end{eqnarray}

\noindent In order to check the gauge invariance, we need to prove
that

\noindent
\begin{equation}
\left(\delta_{\xi}^{\left(0\right)}+\delta_{\xi}^{\left(1\right)}\right)\left(S^{\left(0\right)}+S^{\left(1\right)}+S^{\left(2\right)}\right)=0.
\end{equation}

\noindent In other words, this equation requires the following conditions

\noindent
\begin{eqnarray}
\delta_{\xi}^{\left(0\right)}S^{\left(0\right)} & = & 0,\label{eq:C1}\\
\delta_{\xi}^{\left(1\right)}S^{\left(2\right)} & = & 0,\label{eq:C2}\\
\delta_{\xi}^{\left(0\right)}S^{\left(1\right)}+\delta_{\xi}^{\left(1\right)}S^{\left(0\right)} & = & 0,\label{eq:C3}\\
\delta_{\xi}^{\left(1\right)}S^{\left(1\right)}+\delta_{\xi}^{\left(0\right)}S^{\left(2\right)} & = & 0.\label{eq:C4}
\end{eqnarray}

\noindent Since  equations (\ref{eq:C1}) and  (\ref{eq:C3})
are T-dual versions of  equations (\ref{eq:C2}) and
(\ref{eq:C4}) respectively, we only need to check  equations (\ref{eq:C1})
and  (\ref{eq:C3}). It is easy to see that  equation
(\ref{eq:C1}) is automatically satisfied since it is the standard gauge transformations of Einstein's gravity.
One only needs to verify equation (\ref{eq:C3}).
Furthermore, since the Lie derivative terms can be combined into total derivatives, one does not need to consider them in the calculation.
Because of the independence of gauge parameters $\xi_{i}$ and $\tilde{\xi}_{i}$,
we can check the gauge invariance with each of them respectively.
For example, to check gauge invariance of equation (\ref{eq:C3}),
we can set $\tilde{\xi}_{i}=0$ and $\xi_{i}\not=0$, and vice versa.

In the language of $O\left(D,D\right)$ symmetry, all gauge transformations
above can be rewritten in terms of $O\left(D,D\right)$ indices. For
a tensor $A_{M}^{\quad N}$, the generalized
Lie derivative is defined as

\begin{equation}
\hat{\mathcal{L}}_{\xi}A_{M}^{\quad N}\equiv\xi^{P}\partial_{P}A_{M}^{\quad N}+\left(\partial_{M}\xi^{P}-\partial^{P}\xi_{M}\right)A_{P}^{\quad N}+\left(\partial^{N}\xi_{P}-\partial_{P}\xi^{N}\right)A_{M}^{\quad P}.
\end{equation}

\noindent The generalized Lie derivative also satisfies the Leibniz
rule. The gauge transformation for the generalized metric is

\noindent
\begin{equation}
\delta_{\xi}\mathcal{H}^{MN}=\hat{\mathcal{L}}_{\xi}\mathcal{H}^{MN}.
\end{equation}

\noindent For a scalar
$S$ or a generalized scalar $A_{M}^{\quad M}$, the generalized Lie derivative is simply

\noindent
\begin{equation}
\hat{\mathcal{L}}_{\xi}S=\xi^{P}\partial_{P}S,\qquad\hat{\mathcal{L}}_{\xi}A_{M}^{\quad M}=\xi^{P}\partial_{P}A_{M}^{\quad M}.
\end{equation}

\noindent To consider the gauge transformations of
any object $W$, we can split it into two parts

\noindent
\begin{equation}
\delta_{\xi}W=\hat{\mathcal{L}}_{\xi}W+\triangle_{\xi}W,\label{eq:gauge of W}
\end{equation}

\noindent where $\triangle_{\xi}$ also satisfies the Leibniz rule:
$\triangle_{\xi}\left(WV\right)=\left(\triangle_{\xi}W\right)V+W\left(\triangle_{\xi}V\right)$.
The first term of (\ref{eq:gauge of W}) is  a Lie derivative,
therefore it is covariant and we do not need to consider it as we explained above.
The second term of (\ref{eq:gauge of W})
transforms as a tensor. Therefore, it suffices to check $\triangle_{\xi}W=0$ to confirm the gauge invariance of the DFT action.
Now, recall the DFT action,

\noindent
\begin{equation}
S=\int dxd\tilde{x}e^{-2d}\mathcal{R},
\label{C2Action}
\end{equation}

\noindent where
\begin{eqnarray}
\mathcal{R} & = & \frac{1}{8}\mathcal{H}^{MN}\partial_{M}\mathcal{H}^{KL}\partial_{N}\mathcal{H}_{KL}-\frac{1}{2}\mathcal{H}^{MN}\partial_{N}\mathcal{H}^{KL}\partial_{L}\mathcal{H}_{MK}\nonumber \\
 &  & -\partial_{M}d\partial_{N}\mathcal{H}^{MN}+4\mathcal{H}^{MN}\partial_{M}d\partial_{N}d.
\end{eqnarray}

\noindent We can find that

\noindent
\begin{equation}
\delta_{\xi}\mathcal{R}=\hat{\mathcal{L}}_{\xi}\mathcal{R}=\xi^{M}\partial_{M}\mathcal{R},
\end{equation}

\noindent with $\triangle_{\xi}\mathcal{R}=0$. Therefore, $\mathcal{R}$
is a gauge scalar. Moreover, the dilaton term gives

\noindent
\begin{equation}
\delta_{\xi}e^{-2d}=\hat{\mathcal{L}}_{\xi}e^{-2d}=\partial_{M}\left(\xi^{M}e^{-2d}\right),
\end{equation}

\noindent where $\hat{\mathcal{L}}_{\xi}e^{-2d}=-2\left(\hat{\mathcal{L}}_{\xi}d\right)e^{-2d}$
and $\hat{\mathcal{L}}_{\xi}d=\xi^{M}\partial_{M}d-\frac{1}{2}\partial_{M}\xi^{M}$.
Since, the wight of this term equals the unity, it is a scalar density.
To calculate the weight of a density, we can use the method
introduced in refs. \cite{Jeon:2010rw}. We first introduce the semi-covariant derivative,
$\nabla_{C}=\partial_{C}+\Gamma_{C}$:

\noindent
\begin{equation}
\nabla_{C}T_{\omega A_{1}A_{2}\cdots A_{n}}=\partial_{C}T_{\omega A_{1}A_{2}\cdots A_{n}}-\omega\Gamma_{\quad BC}^{B}T_{\omega A_{1}A_{2}\cdots A_{n}}+\sum_{i=1}^{n}\Gamma_{CA_{i}}^{\quad\quad B}T_{\omega A_{1}\cdots A_{i-1}BA_{i+1}\cdots A_{n}},
\end{equation}

\noindent where $T_{\omega A_{1}A_{2}\cdots A_{n}}$ is a field and
$\omega$ is the weight to identify each field. For example, considering
the dilaton term $e^{-2d}$, it is easy to find

\noindent
\begin{equation}
\nabla_{C}e^{-2d}=\left(-2\nabla_{C}d\right)e^{-2d}=\partial_{C}e^{-2d}-\Gamma_{\quad BC}^{B}e^{-2d},
\end{equation}

\noindent where $\nabla_{C}d=\partial_{C}d+\frac{1}{2}\Gamma_{\quad BC}^{B}$.
It implies that the dilaton potential $e^{-2d}$ has a weight $\omega=1$.

In summary, the action is gauge invariant under the strong constraint. However,
if we introduce  a dilaton potential, say, $V\left(d\right)=e^{8d}$, we will
get a term

\begin{equation}
\int dxd\tilde{x}e^{-2d}V\left(d\right)=\int dxd\tilde{x}e^{6d}.
\end{equation}

\noindent The weight of this term is not the unity, thus
it is not a scalar density and breaks the gauge invariance. In the
next section, we will solve this problem by redefining the dilaton, a generalisation of the results in the traditional string cosmology \cite{Gasperini:2003pb}.

\section{Non-local dilaton potential in DFT}
\noindent We would like to emphasize again that the strong constraint $\partial^{M}\partial_{M}\left(\cdot\right)$
is necessary to make the DFT action (\ref{C2Action}) gauge invariant.  Therefore, we are not trying to construct a gauge invariant potential without imposing the strong constraint, though it must be $O(D,D)$ invariant at the first place. We define a non-local $O(D,D)$ invariant  dilaton $d_* (x,\tilde x)$ as

\begin{eqnarray}
e^{-2d_{*}\left(x,\tilde{x}\right)} & \equiv & \int d^{D}x^{\prime}d^{D}\tilde{x}^{\prime}e^{-2d\left(x^{\prime},\tilde{x}^{\prime}\right)}\left[2\sqrt{-\partial_{\mu}\phi\left(x^{\prime},\tilde{x}^{\prime}\right)\partial^{\mu}\phi\left(x^{\prime},\tilde{x}^{\prime}\right)}\delta\left(\phi\left(x,\tilde{x}\right)-\phi\left(x^{\prime},\tilde{x}^{\prime}\right)\right)\right.\nonumber \\
 &  & \left.+2\sqrt{-\tilde{\partial}_{\mu}\tilde{\phi}\left(x^{\prime},\tilde{x}^{\prime}\right)\tilde{\partial}^{\mu}\tilde{\phi}\left(x^{\prime},\tilde{x}^{\prime}\right)}\delta\left(\tilde{\phi}\left(x,\tilde{x}\right)-\tilde{\phi}\left(x^{\prime},\tilde{x}^{\prime}\right)\right)\right],\label{eq:DFT non dila}
\end{eqnarray}

\noindent where $\tilde g_{\mu\nu} = g^{\mu\nu}$, $d(x',\tilde x')= \tilde{\phi}+\frac{1}{4}\ln\left(-g\right)=\phi
-\frac{1}{4}\ln\left(-g\right)$ to preserve the $O\left(D,D\right)$ symmetry. To check the gauge invariance of this non-local dilaton $d_* (x,\tilde x)$ under the strong constraint, we first write it in the form
\begin{equation}
d_*(x,\tilde x) = \mathcal{V}_{*}^{\left(0\right)} (x,\tilde x)+\mathcal{V}_{*}^{\left(2\right)} (x,\tilde x),
\label{eq:poten action}
\end{equation}

\noindent where the superscript is the number of $\tilde{\partial}$
derivatives and $\mathcal{V}_{*}^{\left(0\right)}$, $\mathcal{V}_{*}^{\left(2\right)}$
are T-dual with each other,

\begin{eqnarray}
\mathcal{V}_{*}^{\left(0\right)} (x,\tilde x) &\equiv& \int d^{D}x^{\prime}d^{D}\tilde{x}^{\prime}e^{-2d\left(x^{\prime},\tilde{x}^{\prime}\right)}2\sqrt{-\partial_{\mu}\phi\left(x^{\prime},\tilde{x}^{\prime}\right)\partial^{\mu}\phi\left(x^{\prime},\tilde{x}^{\prime}\right)}\,\delta\left(\phi\left(x,\tilde{x}\right)-\phi\left(x^{\prime},\tilde{x}^{\prime}\right)\right),\nonumber\\
\mathcal{V}_{*}^{\left(2\right)} (x,\tilde x) &\equiv& \int d^{D}x^{\prime}d^{D}\tilde{x}^{\prime}e^{-2d\left(x^{\prime},\tilde{x}^{\prime}\right)}2\sqrt{-\tilde{\partial}_{\mu}\tilde{\phi}\left(x^{\prime},\tilde{x}^{\prime}\right)\tilde{\partial}^{\mu}\tilde{\phi}\left(x^{\prime},\tilde{x}^{\prime}\right)}\,\delta\left(\tilde{\phi}\left(x,\tilde{x}\right)-\tilde{\phi}\left(x^{\prime},\tilde{x}^{\prime}\right)\right).\nonumber\\
& & \label{eq:individual V definition}
\end{eqnarray}

\noindent It has no harm to include a $\mathcal{V}_{*}^{\left(1\right)}$ term in the definition (\ref{eq:DFT non dila}), nevertheless it will be killed by the strong constraint. To respect the gauge symmetry, under the strong constraint, equation (\ref{eq:poten action}) must satisfy

\noindent
\begin{equation}
\left(\delta_{\xi}^{\left(0\right)}+\delta_{\xi}^{\left(1\right)}\right) \left(\mathcal{V}_{*}^{\left(0\right)}+\mathcal{V}_{*}^{\left(2\right)}\right)=0.
\end{equation}

\noindent We know that $\delta_{\xi}^{\left(0\right)}\mathcal{V}_{*}^{\left(0\right)}=0$
is the standard gauge invariance of the traditional string cosmology,
and $\delta_{\xi}^{\left(1\right)}\mathcal{V}_{*}^{\left(2\right)}=0$
is the T-dual version of it. Since $\delta_{\xi}^{\left(0\right)}\mathcal{V}_{*}^{\left(2\right)}$ is the T-dual of $\delta_{\xi}^{\left(1\right)}\mathcal{V}_{*}^{\left(0\right)}$, we only need to check  $\delta_{\xi}^{\left(1\right)}\mathcal{V}_{*}^{\left(0\right)}=0$
by setting $\xi_{i}$ non-zero or $\tilde{\xi}_{i}$ non-zero
respectively. Bear in mind that imposing the strong constraint
is equivalent to setting all fields having dependence on only half of the doubled coordinates. Even the first term in equation (\ref{eq:DFT non dila}), the original DFT action, is not gauge invariant without imposing the strong constraint.
Therefore, when check the gauge invariance of $d_*(x,\tilde x)$, we can assume all field to depend on only one set of coordinates. Looking back at equation (\ref{eq:individual V definition}), it implies that to get a nonvanishing $\mathcal{V}_{*}^{\left(0\right)}$, we have $\tilde\partial \cdot A =0$ for an arbitrary field or parameter $A$.

%
%
%

When $\tilde{\xi}_{i}$ is non-zero, we
have

\begin{equation}
\delta^{\left(1\right)}g_{ij}=\mathcal{L}_{\tilde{\xi}}g_{ij},\qquad\delta^{\left(1\right)}b_{ij}=\mathcal{L}_{\tilde{\xi}}b_{ij},\qquad\delta^{\left(1\right)}d=\tilde{\xi}_{i}\tilde{\partial}^{i}d-\frac{1}{2}\tilde{\partial}^{i}\tilde{\xi}_{i}.
\end{equation}

\noindent We thus obtain

\begin{eqnarray}
\delta^{\left(1\right)}\mathcal{V}_{*}^{\left(0\right)}
 & = & \int d^{D}x^{\prime}d^{D}\tilde{x}^{\prime}\tilde{\partial}^{i}\left(\tilde{\xi}_{i}e^{-2d}\right)2\sqrt{-\partial_{\mu}\phi\left(x^{\prime},\tilde{x}^{\prime}\right)\partial^{\mu}\phi\left(x^{\prime},\tilde{x}^{\prime}\right)}\delta\left(\phi\left(x,\tilde{x}\right)-\phi\left(x^{\prime},\tilde{x}^{\prime}\right)\right)\nonumber \\
 & + & \int d^{D}x^{\prime}d^{D}\tilde{x}^{\prime}e^{-2d\left(x^{\prime},\tilde{x}^{\prime}\right)}2\left(\delta^{\left(1\right)}\sqrt{-\partial_{\mu}\phi\left(x^{\prime},\tilde{x}^{\prime}\right)\partial^{\mu}\phi\left(x^{\prime},\tilde{x}^{\prime}\right)}\right)\delta\left(\phi\left(x,\tilde{x}\right)-\phi\left(x^{\prime},\tilde{x}^{\prime}\right)\right),\nonumber \\ &  &
\end{eqnarray}

\noindent where we used $\delta^{\left(1\right)}e^{-2d}=\tilde{\partial}^{i}\left(\tilde{\xi}_{i}e^{-2d}\right)$ to get the first term on the r.h.s. Since under the strong constraint,   $\delta^{\left(1\right)}\phi= \tilde{\xi}_{i}\tilde{\partial}^{i} \phi+\tilde{\partial}^{i}\tilde{\xi}_{i}=0$, the second term on the r.h.s vanishes and the first term is a total derivative. Therefore, we have $\delta^{\left(1\right)} \mathcal{V}_{*}^{\left(0\right)}=0$. On the other hand, when $\xi_{i}$ is non-zero, the gauge variations are

\noindent
\begin{align}
\delta^{\left(1\right)}g_{ij} & =2\left(\tilde{\partial}^{k}\xi^{l}-\tilde{\partial}^{l}\xi^{k}\right)g_{k\left(i\right.}b_{\left.j\right)l},\nonumber \\
\delta^{\left(1\right)}g^{ij} & =-\left(\tilde{\partial}^{i}\xi^{k}-\tilde{\partial}^{k}\xi^{i}\right)g^{jl}b_{lk}+\left(i\leftrightarrow j\right),\nonumber \\
\delta^{\left(1\right)}b_{ij} & =g_{ik}\left(\tilde{\partial}^{l}\xi^{k}-\tilde{\partial}^{k}\xi^{l}\right)g_{lj}+b_{ik}\left(\tilde{\partial}^{l}\xi^{k}-\tilde{\partial}^{k}\xi^{l}\right)b_{lj},
\end{align}

\noindent and we have

\begin{equation}
\delta^{\left(1\right)}\mathcal{V}_{*}^{\left(0\right)}=\delta^{\left(1\right)}\int d^{D}x^{\prime}d^{D}\tilde{x}^{\prime}e^{-2d\left(x^{\prime},\tilde{x}^{\prime}\right)}
2\sqrt{-\partial_{\mu}\phi\left(x^{\prime},\tilde{x}^{\prime}\right)\partial^{\mu}
\phi\left(x^{\prime},\tilde{x}^{\prime}\right)}\delta\left(\phi\left(x,\tilde{x}\right)
-\phi\left(x^{\prime},\tilde{x}^{\prime}\right)\right).
\end{equation}

\noindent Since  $\tilde{\xi}_{i}=0$, we have $\delta^{\left(1\right)}d=\tilde{\xi}_{i}\tilde{\partial}^{i}d-\frac{1}{2}\tilde{\partial}^{i}\tilde{\xi}_{i}=0$. After applying the strong constraint, it is easy to see  $\delta^{\left(1\right)}\phi=\tilde{\xi}_{i} \tilde{\partial}^{i}\phi+\tilde{\partial}^{i}\tilde{\xi}_{i}=0$. We thus conclude $\delta^{\left(1\right)} \mathcal{V}_{*}^{\left(0\right)}=0$ for both cases and then $d_* (x,\tilde x)$ is a gauge scalar under the strong constraint. Moreover, since the definition of the non-local dilaton (\ref{eq:DFT non dila}) is independent of $b$-field, we do not need to consider the C-bracket\footnote{We thank C. Ma for reminding us of this.}, and the closure of the Lie algebra is preserved \cite{Hull:2009mi}.

Given $d_* (x,\tilde x)$ is a gauge scalar, phenomenologically, any regular function of $d_* (x,\tilde x)$ could serve as a non-trivial dilaton potential in the DFT action

\begin{equation}
S=\int dxd\tilde{x}e^{-2d}\Big[\mathcal{R}-V\big(d_* (x,\tilde x)\big)\Big].\label{eq:modi action}
\end{equation}

\noindent Nevertheless, it is possible to derive a more realistic dilaton potential from loop corrections. To fulfill this purpose, one needs first to show that in DFT,  the $n$-th loop correction $S_n$ can be organized in the form

\begin{equation}
S_n=\int dxd\tilde{x}e^{-2d}e^{2nd_*}(\cdots), \hspace{7mm} n\ge 1,
\end{equation}
where $(\cdots)$ denotes gauge and $O(D,D)$ scalars. Then, loop by loop, dilaton potentials can be obtained by solving the equations of motion.

We claim that if the fields have only one single coordinate dependence, the non-local dilaton (\ref{eq:DFT non dila}) reduces to the ordinary one. Let us choose the cosmological background with the cosmic-time gauge ($g_{00}=-1$), the non-local dilaton (\ref{eq:DFT non dila}) becomes

\begin{equation}
d_{*}\left(t\right)=-\frac{1}{2}\ln V_{d}\int d\phi\left(t^{\prime}\right)\sqrt{-g\left(t^{\prime}\right)}e^{-2\phi\left(t^{\prime}\right)}\delta\left(\phi\left(t\right)-\phi\left(t^{\prime}\right)\right)=d\left(t\right)-\frac{1}{2}\ln V_{d},
\end{equation}

\noindent or

\begin{equation}
d_{*}\left(\tilde{t}\right)=-\frac{1}{2}\ln\tilde{V}_{d}\int d\tilde{\phi}\left(\tilde{t}^{\prime}\right)\sqrt{-\tilde{g}\left(\tilde{t}^{\prime}\right)}e^{-2\tilde{\phi}\left(\tilde{t}^{\prime}\right)}\delta\left(\tilde{\phi}\left(\tilde{t}\right)-\tilde{\phi}\left(\tilde{t}^{\prime}\right)\right)=d\left(\tilde{t}\right)-\frac{1}{2}\ln\tilde{V}_{d},
\end{equation}

\noindent where $V_{d}=\int d^{d}x^{\prime}d^{D}\tilde{x}^{\prime}$ and
$\tilde{V}_{d}=\int d^{D}x^{\prime}d^{d}\tilde{x}^{\prime}$.
In the previous work \cite{Wu:2013ixa}, we have shown that the DFT action
can be simplified in the cosmological background:

\begin{equation}
S=-\int dtd\tilde{t}e^{-2d}\left[\frac{1}{8}\mathrm{Tr}\left(\frac{\partial M}{\partial\tilde{t}}\frac{\partial M^{-1}}{\partial\tilde{t}}\right)+4\left(\frac{\partial d}{\partial\tilde{t}}\right)^{2}+\frac{1}{8}\mathrm{Tr}\left(\frac{\partial M}{\partial t}\frac{\partial M^{-1}}{\partial t}\right)+4\left(\frac{\partial d}{\partial t}\right)^{2}\right],
\end{equation}

\noindent with

\begin{equation}
M=\left(\begin{array}{cc}
G^{-1} & -G^{-1}B\\
BG^{-1} & G-BG^{-1}B
\end{array}\right),\label{M}
\end{equation}

\noindent where $G$ and $B$ are spatial parts of $g_{ij}\left(t\right)$
and $b_{ij}\left(t\right)$.
To obtain regular solutions which smoothly connect the pre- and post-
big bangs,
we set the dilaton potential to take a special form

\begin{equation}
V\left(t,\tilde{t}\right)=V_{0}e^{8d_{*}\left(t,\tilde{t}\right)}.\label{Cos dila pote}
\end{equation}

\noindent where $V_{0}$ is a function of $V_{d}$ and $\tilde{V}_{d}$,
With this non-local dilaton potential, we find

\begin{eqnarray}
S & = & -\int dtd\tilde{t}e^{-2d}\left[\frac{1}{8}\mathrm{Tr}\left(\frac{\partial M}{\partial\tilde{t}}\frac{\partial M^{-1}}{\partial\tilde{t}}\right)+4\left(\frac{\partial d}{\partial\tilde{t}}\right)^{2}\right.\nonumber \\
 &  & \left.+\frac{1}{8}\mathrm{Tr}\left(\frac{\partial M}{\partial t}\frac{\partial M^{-1}}{\partial t}\right)+4\left(\frac{\partial d}{\partial t}\right)^{2}-V_{0}e^{8d_{*}\left(t,\tilde{t}\right)}\right].
\end{eqnarray}

\noindent To calculate the EOM of this action, the strong constraint oughts to be imposed and then all fields depend on only one temporal direction. Therefore, the dilaton and the redefined non-local dilaton coincide as $d\left(t\right)$ or $d\left(\tilde{t}\right)$.
Two optional actions are given as follows:

\begin{equation}
S=-\int dtd\tilde{t}e^{-2d}\left[\frac{1}{8}\mathrm{Tr}\left(\frac{\partial M}{\partial\tilde{t}}\frac{\partial M^{-1}}{\partial\tilde{t}}\right)+4\left(\frac{\partial d}{\partial\tilde{t}}\right)^{2}-V_{0}e^{8d}\right],
\end{equation}

\noindent or

\begin{equation}
S=-\int dtd\tilde{t}e^{-2d}\left[\frac{1}{8}\mathrm{Tr}\left(\frac{\partial M}{\partial t}\frac{\partial M^{-1}}{\partial t}\right)+4\left(\frac{\partial d}{\partial t}\right)^{2}-V_{0}e^{8d}\right].
\end{equation}

\noindent The solutions of these two actions and their physical implications
are given in refs. \cite{Wu:2013sha,Wu:2013ixa}.

\section{Conclusion}

In literature, it was believed that DFT only admits trivial dilaton potentials.  In this paper, after presenting the gauge
transformations of DFT, we introduced a non-local $O(D,D)$ invariant
dilaton in the DFT formalism. We showed that this
non-local dilaton is a consistent gauge invariant under the strong constraint. It is therefore possible to include loop corrections in the formalism of DFT.  Our construction reduces to the ordinary dilaton when the fields depend on one single coordinate. It is of interest to extend the construction to more general backgrounds. The strong constraint is crucial in our construction. There may exist some weaker conditional results.
Moreover, it is also significant to consider the dilaton potential combined
with $\alpha^{\prime}$ corrections, where the gauge transformations are slightly modified.

\acknowledgments

This work is supported in part by the NSFC (Grant No. 11175039 and 11375121
) and SiChuan Province Science Foundation for Youths (Grant No. 2012JQ0039).



\begin{thebibliography}{99}

\bibitem{Siegel:1993xq}
  W.~Siegel,
  ``Two vierbein formalism for string inspired axionic gravity,''
  Phys.\ Rev.\ D {\bf 47}, 5453 (1993)
  [hep-th/9302036].
  W.~Siegel,
  ``Superspace duality in low-energy superstrings,''
  Phys.\ Rev.\ D {\bf 48}, 2826 (1993)
  [hep-th/9305073].
  W.~Siegel,
  ``Manifest duality in low-energy superstrings,''
  In *Berkeley 1993, Proceedings, Strings '93* 353-363, and State U. New York Stony Brook - ITP-SB-93-050 (93,rec.Sep.) 11 p. (315661)
  [hep-th/9308133].


\bibitem{Hull:2009mi}
C.~Hull and B.~Zwiebach,   ``Double Field Theory,''   JHEP {\bf 0909}, 099 (2009)   [arXiv:0904.4664 [hep-th]].   
C.~Hull and B.~Zwiebach,   ``The Gauge algebra of double field theory and Courant brackets,''  JHEP {\bf 0909}, 090 (2009)  [arXiv:0908.1792 [hep-th]].  
O.~Hohm, C.~Hull and B.~Zwiebach,   ``Background independent action for double field theory,''   JHEP {\bf 1007}, 016 (2010)   [arXiv:1003.5027 [hep-th]].   
O.~Hohm, C.~Hull and B.~Zwiebach,   ``Generalized metric formulation of double field theory,''   JHEP {\bf 1008}, 008 (2010)   [arXiv:1006.4823 [hep-th]].   


\bibitem{Tseytlin:1990nb}   A.~A.~Tseytlin,   ``Duality Symmetric Formulation Of String World Sheet Dynamics,''   Phys.\ Lett.\ B {\bf 242}, 163 (1990).   
A.~A.~Tseytlin,   ``Duality symmetric closed string theory and interacting chiral scalars,''   Nucl.\ Phys.\ B {\bf 350}, 395 (1991).   




\bibitem{Duff:1989tf}   M.~J.~Duff,   ``Duality Rotations In String Theory,''   Nucl.\ Phys.\ B {\bf 335}, 610 (1990).   
M.~J.~Duff and J.~X.~Lu,   ``Duality Rotations In Membrane Theory,''   Nucl.\ Phys.\ B {\bf 347}, 394 (1990).   






\bibitem{Zwiebach:2011rg}   B.~Zwiebach,   ``Double Field Theory, T-Duality, and Courant Brackets,''   arXiv:1109.1782 [hep-th].   

\bibitem{Aldazabal:2013sca}
G.~Aldazabal, D.~Marques and C.~Nunez,   ``Double Field Theory: A Pedagogical Review,''   arXiv:1305.1907 [hep-th].   

\bibitem{Berman:2013eva}
D.~S.~Berman and D.~C.~Thompson,   ``Duality Symmetric String and M-Theory,''  arXiv:1306.2643 [hep-th].  

\bibitem{Hohm:2013bwa}
  O.~Hohm, D.~Lust and B.~Zwiebach,
  ``The Spacetime of Double Field Theory: Review, Remarks, and Outlook,''  arXiv:1309.2977 [hep-th].  


\bibitem{Hohm:2011cp} O.~Hohm and S.~K.~Kwak,   ``Massive Type II in Double Field Theory,''  JHEP {\bf 1111}, 086 (2011)  [arXiv:1108.4937 [hep-th]].  


\bibitem{Aldazabal:2011nj} G.~Aldazabal, W.~Baron, D.~Marques and C.~Nunez,   ``The effective action of Double Field Theory,''  JHEP {\bf 1111}, 052 (2011)  [Erratum-ibid.\  {\bf 1111}, 109 (2011)]  [arXiv:1109.0290 [hep-th]].  


\bibitem{Geissbuhler:2011mx} D.~Geissbuhler,   ``Double Field Theory and N=4 Gauged Supergravity,''  JHEP {\bf 1111}, 116 (2011)  [arXiv:1109.4280 [hep-th]].  


\bibitem{Geissbuhler:2013uka}    D.~Geissbuhler, D.~Marques, C.~Nunez and V.~Penas,   ``Exploring Double Field Theory,''   arXiv:1304.1472 [hep-th].   

\bibitem{Berman:2013uda}
  D.~S.~Berman, C.~D.~A.~Blair, E.~Malek and M.~J.~Perry,
  ``The $O_{D,D}$ geometry of string theory,''
  Int.\ J.\ Mod.\ Phys.\ A {\bf 29}, no. 15, 1450080 (2014)
  [arXiv:1303.6727 [hep-th]].

\bibitem{Grana:2012rr} M.~Grana and D.~Marques,   ``Gauged Double Field Theory,''  JHEP {\bf 1204}, 020 (2012)  [arXiv:1201.2924 [hep-th]].  

\bibitem{Jeon:2011cn}
  O.~Hohm and S.~K.~Kwak,
  ``Frame-like Geometry of Double Field Theory,''
  J.\ Phys.\ A {\bf 44}, 085404 (2011)
  [arXiv:1011.4101 [hep-th]].
  I.~Jeon, K.~Lee and J.~-H.~Park,
  ``Stringy differential geometry, beyond Riemann,''
  Phys.\ Rev.\ D {\bf 84}, 044022 (2011)
  [arXiv:1105.6294 [hep-th]].
  O.~Hohm and B.~Zwiebach,
  ``On the Riemann Tensor in Double Field Theory,''
  JHEP {\bf 1205}, 126 (2012)
  [arXiv:1112.5296 [hep-th]].
  O.~Hohm and B.~Zwiebach,
  ``Towards an invariant geometry of double field theory,''
  J.\ Math.\ Phys.\  {\bf 54}, 032303 (2013)
  [arXiv:1212.1736 [hep-th]].

\bibitem{Gasperini:2007zz}
M.~Gasperini,   ``Elements of string cosmology,''   Cambridge, UK: Cambridge Univ. Pr. (2007) 552 p.


\bibitem{Hohm:2013jaa}
  O.~Hohm, W.~Siegel and B.~Zwiebach,
  ``Doubled $\alpha'$-geometry,''
  JHEP {\bf 1402}, 065 (2014)
  [arXiv:1306.2970 [hep-th]].
  O.~Hohm and B.~Zwiebach,
  ``Green-Schwarz mechanism and $\alpha'$-deformed Courant brackets,''
  arXiv:1407.0708 [hep-th].
  O.~Hohm and B.~Zwiebach,
  ``Double Field Theory at Order $\alpha'$,''
  arXiv:1407.3803 [hep-th].
  O.~A.~Bedoya, D.~Marques and C.~Nunez,
  ``Heterotic $\alpha$'-corrections in Double Field Theory,''
  arXiv:1407.0365 [hep-th].





\bibitem{Gasperini:2003pb}
  M.~Gasperini, M.~Giovannini and G.~Veneziano,
  ``Perturbations in a nonsingular bouncing universe,''
  Phys.\ Lett.\ B {\bf 569}, 113 (2003)
  [hep-th/0306113].

\bibitem{Wu:2013sha}
  H.~Wu and H.~Yang,
  ``Double Field Theory Inspired Cosmology,''
  JCAP {\bf 1407}, 024 (2014)
  [arXiv:1307.0159 [hep-th]].


\bibitem{Wu:2013ixa}
  H.~Wu and H.~Yang,
  ``New Cosmological Signatures from Double Field Theory,''
  arXiv:1312.5580 [hep-th].




\bibitem{Ma:2014ala}
  C.~-T.~Ma and C.~-M.~Shen,
  ``Cosmological Implications from O(D,D),''
  arXiv:1405.4073 [hep-th].





\bibitem{Jeon:2010rw} I.~Jeon, K.~Lee and J.~-H.~Park,   ``Differential geometry with a projection: Application to double field theory,''  JHEP {\bf 1104}, 014 (2011)  [arXiv:1011.1324 [hep-th]].  
I.~Jeon, K.~Lee and J.~-H.~Park,   ``Double field formulation of Yang-Mills theory,''  Phys.\ Lett.\ B {\bf 701}, 260 (2011)  [arXiv:1102.0419 [hep-th]].  
J.~-H.~Park,   ``Comments on double field theory and diffeomorphism,''   arXiv:1304.5946 [hep-th].   
  O.~Hohm and B.~Zwiebach,
  ``Large Gauge Transformations in Double Field Theory,''
  JHEP {\bf 1302}, 075 (2013)
  [arXiv:1207.4198 [hep-th]].









\end{thebibliography}
\end{document}